\documentclass[english,twoside,a4paper,10pt]{article}

\usepackage[latin1]{inputenc}
\usepackage[T1]{fontenc}
\usepackage{amsmath}
\usepackage{amsfonts}
\usepackage{graphicx}
\usepackage{subfigure}
\usepackage{a4wide}
\usepackage{amssymb}
\usepackage{fancyhdr}
\usepackage{mathrsfs}
\usepackage[toc,page]{appendix}
\usepackage{xcolor}

\def\ba{\begin{eqnarray}}
\def\ea{\end{eqnarray}}

\def\be{\begin{equation}}
\def\ee{\end{equation}}

\begin{document}

\title{First law of thermodynamics and entropy of FLRW universe in modified gravity}

\author{ 
Lorenzo Sebastiani \footnote{E-mail address:lorenzo.sebastiani@unitn.it}
\\ \\
\begin{small}
Dipartimento di Fisica, Universit\`a di Trento, Via Sommarive 14, 38123 Povo (TN), Italy
\end{small}\\
}

\date{}

\maketitle

\abstract{We investigate the first law of thermodynamics and entropy associated to the apparent horizon of (non-flat) FLRW space-time in different theories of modified gravity and in the presence of a perfect fluid of matter. We pose our attention on those theories which lead to second order differential field equations on FLRW background. In this way,  we observe that one may obtain a formula for entropy in terms of the radius of the apparent horizon only. Thus, when considering a modification to the area law of General Relativity, it is possible to reconstruct the gravitational lagrangian consistent with the corresponding first law.}

\section{Introduction}

In General Relativity (GR), several thermodynamical quantities (energy, surface gravity, temperature,  entropy...) may be introduced for black holes
by using semiclassical approaches based on quantum mechanical methods in curved space-times. In particular, the Hawking radiation
\cite{Hawrad1, Hawrad2,Hawrad3,Hawrad4,Hawrad5, Visser}
that takes place on the black hole event horizon implies that black holes have temperature $T_H$ and the first law of thermodynamics holds true in the form $dE=T_H dS$, once 
the notion of quasi-local
Misner-Sharp gravitational energy $E$
is assumed. Thus,
the first law is consistent with the Bekenstein-Hawking entropy $S=\frac{A_H}{4}$
\cite{entropyBH1, entropyBH2}, also known as the area law, which is proportional to the area $A_H$ of the black hole horizon. Moreover, it has been shown that it is also possible to derive the Einstein's field equations by starting from
the first law of black hole thermodynamics \cite{jacobson}.

The thermodynamical properties of the black hole horizon can be extended to generic space-time horizon and in the specific to the apparent horizon of Friedmann - Lema\^itre - Robertson - Walker (FLRW) space-time describing our observable homogeneous and isotropic universe. 
The temperature associated to the apparent horizon can be inferred from the surface gravity with a covariant formalism, and if one assumes the validity of the entropy area law, it is possible to recast the Friedmann equations of GR in the form of the first law $dE=T_H dS$, where $dE$ is the amount of energy flux crossing the apparent horizon \cite{flux1, flux2, flux3}.
Further generalizations of this result have been investigated for generic
$(n+1)$-dimensional FLRW universe with any spatial curvature,  Gauss-Bonnet and Lovelock gravity \cite{GBL1,GBL2} and for scalar-tensor gravity \cite{scalartensor1, scalartensor2} (see also Refs. \cite{Bousso, Calcagni, Pad1, Pad2}).

However, in FLRW universe a source of perfect fluid is present and, differently from the black hole case, 
there are well-posed concepts of energy density $\rho$ and pressure $p$ of the fluid. Thus, 
in GR
the Einstein's field equations evaluated on the apparent horizon may be also interpreted as a first law with the account of a working term, namely  $dE=WdV_H+T_HdS$, where $E=\rho V_H$ and $W=\frac{(\rho-p)}{2}$, $V_H$ being the volume enclosed by apparent horizon. This procedure can be extended to Gauss-Bonnet and Lovelock gravity \cite{GBworking1, GBworking2} and in all this cases the results lead to the identification of entropy with the area law (see also Refs. \cite{brane1, brane2} for braneworld scenario and Ref. \cite{apparenthorizon} for the issues related to  thermodynamics of apparent horizon).

On the other hand, when one moves to the framework of modified theories of gravity, the field equations become quite involved and the area law  is not still valid (see for example Refs. \cite{Odentropy, Faraoni}). Therefore, the derivation of a first law from the field equations may furnish a way to define the entropy. 
In Ref. \cite{SuperCai} the case of $F(R)$-gravity, where the gravitational action is expressed by a function $F(R)$ of the Ricci scalar $R$, has been investigated and it has been shown that the first law on FLRW space-time brings to some additional terms to the expected Wald entropy. This result may be due to non-equilibrium thermodynamics of space-time \cite{jacobson2, noneq2}. 

Here, we should mention that since the area law of Bekenstein-Hawking is not an extensive measure, several modified entropy laws have been proposed in the last decades for gravitational systems, i.e. Tsallis entropy \cite{Tsallis1, Tsallis2}, R\'enyi entropy \cite{Renyi}, Kaniadakis entropy  \cite{Kan}, logaritmic corrected entropy \cite{logcorr1, logcorr2, logcorr3} and, more recently, Barrow entropy \cite{Barrowen}. In Ref. \cite{Santos} an attempt to rewrite the first law for $F(R)$-gravity by using the Barrow entropy has carried out.

The fact that the Einstein's field equations of GR are a system of partial differential equations which are at most at the second order in the derivatives has profound mathematical and physical implications that cannot be transposed in a modified theory of gravity. The Lovelock's theorem \cite{Love} states that in four dimension the only lagrangian depending on the curvature invariants only and admitting second order differential equations is given by the Hilbert-Einstein action of GR (up to the cosmological constant). However, there are some classes of theories which preserve second order differential equations on four dimensional FLRW space-time, for example
the $F(R, P, Q)$-models, where $P$ is the square of the Ricci tensor and $Q$ is the square of the Riemann tensor, derived in Ref. \cite{Gao} or
the so called non-polynomial models investigated in Ref. \cite{Chinaglia1, Chinaglia2}. In these frameworks, some thermodynamic issue on FLRW space-time is much more tractable. We also note that this models are inspired by Quantum Loop Cosmology (QLC) and their FLRW solutions are singularity free, admitting the bounce as an alternative scenario for early-time universe (see Ref. \cite{Casaz} and reference therein). 

In this paper, we investigate the first law of thermodynamics associated to the apparent horizon of non-flat FLRW metric and in the presence of a perfect fluid by posing our attention on gravitational theories with Friedmann-like second-order differential field equations. The presence of perfect fluid gives rise to a working term in the first law and the entropy is derived consistently with the equations of motion. As a result, at least in the flat spatial case, we found that the entropy can be expressed in terms of the radius of the apparent horizon only and can be computed independently of the explicit form of the scale factor. 
We comment the results in the light of the various entropy scenarios.    

The paper is organized as follows. In Sec.{\bf 2} we introduce the formalism. The case of $F(R, G)$-gravity is investigated in Sec.{\bf 3}, while in Sec.{\bf 4} we analyze two non-polynomial gravity models depending on the covariant derivatives of the Ricci scalar and of the Gauss-Bonnet. Sec.{\bf 5} and Sec.{\bf 6} are devoted to extended mimetic gravity models. Conclusions and final remarks are given in Sec.{\bf 7}.

We use the Newton's gravitational constant $G_N=1$.

\section{Formalism}

We work in a non-flat four dimensional FLRW space-time whose metric is given by,
\begin{equation}
ds^2=-dt^2+a(t)^2\left(\frac{d r^2}{1-k r^2}+ r^2d\Omega^2_2\right)\,,\label{metric}
\end{equation}
where $a\equiv a(t)$ is the scale factor of the universe depending on cosmological time only, $d\Omega_2^2\equiv d\theta^2+\sin^2\theta d\phi^2$ is the metric of a two-dimensional sphere and the parameter $k=0,\pm 1$ corresponds to the spatial curvature.  

We introduce the relevant invariant scalar quantity,
\begin{equation}
\chi=\gamma^{a b}\partial_a r\partial_b r\,.
\end{equation}
Here, $\gamma_{ab}$ is the two-dimensional metric related to $a,b=0,1$. On FLRW space-time we get
\begin{equation}
\chi=1-r^2 J^2\,,\quad\quad J^2=\left(H^2+\frac{k}{a^2}\right)\,,
\label{chi}
\end{equation}
where $H\equiv H(t)=\frac{\dot a}{a}$ is the Hubble parameter, the dot being the time derivative.
We take as the physical boundary of the universe the apparent horizon whose radius $r_H$ is given by
\begin{equation}
\chi=0\quad\quad\rightarrow \quad\quad
r_H= \frac{1}{\sqrt{H^2+k/a^2}}\,.   \label{rH}
\end{equation}
Now it is possible to associate to the apparent horizon a temperature $T_H$ as,
\begin{equation}
T_H\equiv\frac{\kappa_H}{2\pi}=\frac{1}{2\pi r_H} \left(-1+\frac{\dot r_H}{2 H r_H}\right)\,,
\label{TH}
\end{equation}
where $\kappa_H$ is the Hayward surface gravity \cite{flux1}. Note that the physical temperature must be assumed to be $T_H=|\kappa_H|/(2\pi)$ in order to avoid negative values.  

The work density is defined as \cite{flux2},
\begin{equation}
W\equiv-\frac{1}{2}T^{ab}\gamma_{ab}=\frac{1}{2}\left(\rho-p\right)\,,\label{W}
\end{equation}
where $T_{\mu\nu}$ is the stress-energy tensor of perfect matter-radiation fluid with energy density $\rho$ and pressure $p$.

Finally, the total amount of energy inside the apparent horizon is given by,
\begin{equation}
E\equiv \rho V_H=\frac{4\pi}{3}\rho r_H^3\,,     
\end{equation}
where $V_H=\frac{4\pi}{3}r_H^3$ is the volume enclosed by apparent horizon.
As a consequence, the differential of energy reads,
\begin{equation}
d E= \frac{4\pi}{3}r_H^3 d\rho+4\pi\rho r_H^2 dr_H\,,
\label{energy}
\end{equation}
and the following relation holds true,
\begin{equation}
dE=\frac{4\pi}{3}r^3_H d\rho+WdV_H+2\pi r_H^2(\rho+p)dr_H\,.\label{dE}
\end{equation}
In the next sections we will consider different modified theories of gravity on FLRW space-time and in the presence of perfect fluid and we will try to recast the equations of motion in the form of a first law of thermodynamics. Therefore, we will infer a formula for the entropy which makes consistent the first law. We will restrict our analysis on specific classes of models which lead to second order differential equations of motion on FLRW space-time.

\section{$F(R, G)$-gravity\label{SecIII}}

Let us start by considering the following model of modified gravity \cite{rev1, rev2, rev3, rev4, rev5, rev6}, 
\begin{eqnarray}
I = \frac{1}{16\pi}\int_\mathcal{M} d^4 x \sqrt{-g}\, \left(F(R, G) \right)+ I_m \,,
\end{eqnarray}
where $g$ is the determinant of the metric tensor, $\mathcal{M}$ is a compact manifold and $I_m$ is the usual action of matter. Here, $F\equiv F(R, G)$ is a function of the Ricci scalar $R$ and
of the Gauss-Bonnet four dimensional topological invariant $G$,
\begin{equation}
G=R^{2}-4R_{\mu\nu}R^{\mu\nu}+R_{\mu\nu\xi\sigma}R^{\mu\nu\xi\sigma}\,,\label{GaussBonnet}
\end{equation}
$R_{\mu\nu}$ and $R_{\mu\nu\xi\sigma}$ being the Ricci tensor and the Riemann tensor, respectively. When $F(R, G)=R$ we recover the Hilbert-Einstein action of GR.
On FLRW space-time (\ref{metric}) we obtain,
\begin{eqnarray}
R&=& 6\left(
\frac{\ddot a}{a }+\frac{\dot a ^2}{a^2 }+\frac{k}{a^2}\right)\equiv \left(12H^2+6\dot H+\frac{6k}{a^2}\right)\,,\label{R}\label{RFLRW}\\
G&=& \frac{24 }{a^3 }\left(
\dot a^2 \ddot a+\ddot a k 
\right)\equiv 24\left(H^2+\frac{k}{a^2}\right)\left(H^2+\dot H\right)\label{G}\,.\label{GFLRW}
\end{eqnarray}
By assuming the matter contents of the universe as a perfect fluid,
the field equations on FLRW can be written as,
\begin{align}
16\pi\rho=&6 \left(H^2+\frac{k}{a^2}\right)F_R+\left(F-RF_R-G F_G\right)+6H\left(\dot F_R+4\left(H^2+\frac{k}{a^2}\right)\dot F_G\right)\,,\label{FRGone}\\
6\pi(\rho+p)=&F_R\left(-4\dot H+\frac{4k}{a^2}\right)
+2H\dot F_R+\dot F_G\left(8H^3+\frac{24Hk}{a^2}-16H\dot H\right)
-2\ddot F_R\nonumber\\&
+\ddot F_G\left(-8H^2-\frac{8k}{a^2}\right)\,,\label{FRGtwo}
\end{align}
where $R$ and $G$ are given by (\ref{RFLRW})--(\ref{GFLRW}) and we use the notation
\begin{equation}
F_R=\frac{\partial F}{\partial R}\,,\quad\quad
F_G=\frac{\partial F}{\partial G}\,.
\end{equation}
In the equations above, $\rho\equiv \rho(t)$ and $p\equiv p(t)$ are the energy density and pressure of the matter-radiation contents of the universe and obey to the conservation law,
\begin{equation}
\dot\rho+3H(\rho+p)=0\,.    \label{conslaw}
\end{equation}
By taking the variation of Eq. (\ref{FRGone}) and by evaluating the result on the apparent horizon $r_H$  (\ref{rH}) we get,
\begin{equation}
\frac{4\pi}{3}r_H^3 d\rho=
-F_R d r_H-\frac{H^2r_H^3}{2}d F_R+\left(
-2H^2 r_H-4H\dot r_H
\right)d F_G+\frac{H r_H^3}{2}d \dot F_R+2H r_H d\dot F_G\,,\label{step1}
\end{equation}
where we have multiplied the both sides of the equation by $r^3/12$ and we have used the following relations:
\begin{equation}
\left(H^2+\frac{k}{a^2}\right)=\frac{1}{r_H^2}\,,\quad \quad 
\left(\dot H-\frac{k}{a^2}\right)=-\frac{\dot r}{H r_H^3}\,.
\end{equation}
Now, from Eq. (\ref{FRGtwo}) we derive
\begin{equation}
2\pi r_H^2(\rho+p)dr_H=
\frac{\dot r_H}{2H r_H}F_R dr_H+\frac{r_H^2 H\dot r_H}{4}d F_R-\frac{r_H^2\dot r_H}{4}
d\dot F_R+H\dot r_H d F_G +\frac{2\dot r_H^2 }{r_H}d F_G-\dot r_H d \dot F_G\,.
\label{step2}
\end{equation}
As a consequence, from Eq. (\ref{step1}) and Eq. (\ref{step2}) together with relation (\ref{dE}) we get
\begin{eqnarray}
dE&=&WdV_H+\left(F_R dr_H+\frac{r_H}{2}dF_R\right)
\left(-1+\frac{\dot r_H}{2 H r_H}\right)
-\frac{H r_H^3}{2}d\dot F_R
\left(-1+\frac{\dot r_H}{2 H r_H}\right)
\nonumber
\\
&&
+\frac{r_H}{2}d F_R\left(H^2 r_H^2 - 1\right)
\left(-1+\frac{\dot r_H}{2 H r_H}\right)+
\frac{2}{r_H}d F_G
\left(-1+\frac{\dot r_H}{2 H r_H}\right)
\nonumber\\
&&+\left(\frac{2}{r_H}(H^2 r_H^2-1)+4H \dot r_H\right)d F_G \left(-1+\frac{\dot r_H}{2 H r_H}\right)
\nonumber\\
&&+\
-2 H r_H d\dot F_G \left(-1+\frac{\dot r_H}{2 H r_H}\right)\,.
\end{eqnarray}
Therefore, by making use of the Wald entropy result for $F(R,G)$-gravity (see Appendix),
\begin{equation}
S_W=\frac{A_H}{4 }\left(F_R+ F_G\left(\frac{4}{r_H^2}\right)\right)\,,
\end{equation}
with the area of the horizon $A_H=4\pi r_H^2$, we can write
\begin{eqnarray}
dE&=&WdV_H+
T_H dS_W+\left(
-\frac{H r_H^3}{2}d\dot F_R
+\frac{r}{2}d F_R\left(H^2 r_H^2 - 1\right)
+\left(\frac{2}{r_H}(H^2 r_H^2-1)+4H \dot r_H\right)d F_G 
\right.
\nonumber\\
&&\left.
-2 H r_H d\dot F_G \right)2\pi r_H T_H\,,
\label{firstlawFRG}
\end{eqnarray}
where we have introduced the horizon temperature (\ref{TH}).
For $F(R)$-gravity only we obtain,
\begin{equation}
dE=Wd V_H+TdS_W+
+\frac{A_H T_H}{4}\left(
-H r_H^2d \dot F_R+H^2 r^2 d F_R- d F_R
\right)\,,
\end{equation}
with $S_W=A_H F_R/4$. This result is in agreement with Refs. \cite{SuperCai, Bamba, Geng}. 

Thus, when the equations of motion of $F(R, G)$-gravity are arranged into a form of the first law of thermodynamics at the apparent horizon, the Wald entropy should be redefined  as $S_W\rightarrow S_W+ \bar S$ such that
\begin{equation}
d \bar S=
2\pi r_H 
\left(
-\frac{H r_H^3}{2}d\dot F_R
+\frac{r}{2}d F_R\left(H^2 r_H^2 - 1\right)
+\left(\frac{2}{r_H}(H^2 r_H^2-1)+4H \dot r_H\right)d F_G 
-2 H r_H d\dot F_G \right)\,.
\label{barS}
\end{equation}
In literature, it has already been observed how this term can be  a consequence of the non-equilibrium thermodynamics within $F(R, G)$-gravity framework \cite{jacobson2} (see the discussion in Ref. \cite{SuperCai} about the case of $F(R)$-gravity).

In general, the theory under consideration is an higher derivative theory,
where  the equations of motion involve the presence of third and fourth order time derivatives of the scale factor $a(t)$. However, there is a suitable choice of the function $F(R,G)$ which makes the field equations (\ref{FRGone})--(\ref{FRGtwo}) at the second order, namely \cite{Gao, Casaz},
\begin{equation}
F(R, G)=R+f(R, G)\,,\quad
\quad f(R, G)=\frac{R+\sqrt{R^2-6G}}{12}\,.
\label{model0}
\end{equation}
It is easy to show that on FLRW metric (\ref{metric}) we get
\begin{equation}
J^2= \left(H^2+\frac{k}{a^2}\right)=\frac{R+\sqrt{R^2-6G}}{12}\,,
\end{equation}
where $J^2$ is the invariant in (\ref{chi}) and on the apparent horizon it reads $J^2=\frac{1}{r_H^2}$.
Note that this choice contains a non analytic dependence on $R$ and
$G$.

Therefore, by
writing $f(R, G)=f(J^2)$,
since
$f_R=-\frac{4}{r^2}f_G$ and $f_G=-\frac{H r^3}{24\dot r} f_{J^2}$, the first law (\ref{firstlawFRG}) reads,
\begin{equation}
d E=W d V_H+T_H  dS_W+\left(-\frac{\pi r_H^4f_{J^2}}{6}
d J^2+\frac{\pi r_H^4}{3}d\left(H^2 f_{J^2}\right)\right)T_H\,,
\end{equation}
where the Wald entropy $S_W$ turns out to be the area law of GR, $S_W=\frac{A_H}{4}$.
The additional term $d\bar S$ (\ref{barS}) is now given by
\begin{equation}
d\bar S=-\frac{\pi r_H^4f_{J^2}}{6}
d J^2+\frac{\pi r_H^4}{3}d\left(H^2 f_{J^2}\right)\,.
\end{equation}
This term can be explicitly computed for a generic FLRW space-time if we assume $k=0$ and $H^2=\frac{1}{r_H^2}$, namely
\begin{equation}
d\bar S=\frac{\pi r_H}{3}
\left(
-f_{H^2} dr_H+r_H d f_{H^2}
\right)\,.
\label{barSk0}
\end{equation}
Thus, given a specific model in the form of (\ref{model0}), $\bar S$ can be derived
via integration of (\ref{barSk0}) independently of the form of the scale factor $a(t)$ and the result only depends on the radius of the apparent horizon. In particular, by making the choice $f(J^2)=\gamma \sqrt{J^2}=\gamma \sqrt{\frac{R+\sqrt{R^2-6G}}{12}}$, $\gamma$ being a constant, $d\bar S=0$ and we recover the first law of General Relativity. In this case, it is easy to see that the modification of gravity disappears from equations of motion on FLRW space-time. 

Furthermore, we are able to reconstruct the
specific gravitational lagrangians associated to some given entropy laws.
For example, three years ago, Barrow proposed an interesting entropy corrected law inspired by the fractal structure of the black hole horizon surface \cite{Barrowen},
\begin{equation}
S_B=\left(\frac{A_H}{4}\right)^{1+\frac{\Delta}{2}}\,,\quad\quad 0\leq \Delta\,.
\label{Barrowentropy}
\end{equation}
The exponent $\Delta$ quantifies quantum deformations and the Bekenstein entropy is recovered when $\Delta=0$.
In general, by assuming the Barrow entropy one may retrieve the gravitational field equations from the first law \cite{Leon}. Here, we can furnish a model of $F(R,G)$-gravity which leads to such field equations in the flat spatial case, since
by using (\ref{barSk0})
with $d\bar S=dS_B-2 \pi r_H dr_H$
one finds that
the model (\ref{model0}) with
\begin{equation}
f(J^2)=\frac{6\pi^{\frac{\Delta}{2}}}{(\Delta-1)}\frac{\left(1+\frac{\Delta}{2}\right)}{\left(1-\frac{\Delta}{2}\right)}\left(J^2\right)^{1-\frac{\Delta}{2}}+6J^2\,,\quad\quad
J^2=\frac{R+\sqrt{R^2-6G}}{4}\,,
\end{equation}
brings to Barrow entropy (\ref{Barrowentropy}). Note that when $\Delta=0$ the function $f(J^2)=0$ and at the leading order of $\Delta$ one has
\begin{equation}
f(J^2)=-3\Delta J^2=-\Delta\left(\frac{R+\sqrt{R^2-6G}}{4}\right)\,.
\end{equation}
In the following sections we still study the first law in some other classes of gravitational models with second order differential equations of motion on FLRW space-time. The cosmological applications of this models have been largely investigated in Ref. \cite{Casaz}.

\section{Non-polynomial gravity models}

Let we now consider the following non polinomial gravity model \cite{NPG1, NPG2},
\be
I =  \frac{1}{8\pi}\int_{\mathcal M} d^4 x \sqrt{-g} \left[\frac{R}{2} + 
\frac{\alpha}{6}\sqrt{(-\nabla_\mu R)^2}\right]
+I_m\,,
\ee
with $\alpha$ dimensional parameter and $I_m$ the matter action. The Friedmann-like equations read,
\begin{equation}
16\pi \rho=6\left(H^2+\frac{k}{a^2}\right)-6\alpha H^3\,,
\label{primanonpol}
\end{equation}
\begin{equation}
16\pi\left(\rho+p\right)=
-4\left(\dot H-\frac{k}{a^2}\right)
+6\alpha H \dot H\,.
\end{equation}
Variation of Eq. (\ref{primanonpol}) on the apparent horizon leads to
\begin{equation}
16\pi d\rho=-\frac{12}{r_H^3}dr_H
-18\alpha H^2 d H\,.
\end{equation}
Thus, by introducing the temperature of the apparent horizon (\ref{TH}) the first law takes the form,
\begin{equation}
dE=WdV_H+T_H\left(2\pi r_H dr_H+3\pi\alpha H^2 r_H^4 dH\right)\,,
\end{equation}
and in the case of $k=0$ we get
\begin{equation}
dE=WdV_H+T_H\left(2\pi r_H dr_H-3\pi\alpha  dr\right)\,,
\end{equation}
such that
\begin{equation}
S=\frac{A_H}{4}\left(1-\frac{3\alpha}{r_H}\right)\,.\label{SNPG1}
\end{equation}
We will come back to this result at the end of the section.\\
\\
A similar example can be found in the following model,
\be
I =  \frac{1}{8\pi}\int_{\mathcal M} d^4 x \sqrt{-g} \left[\frac{R}{2} + 
\frac{\alpha}{6}\sqrt{(-\nabla_\mu G)^2}\right]
+I_m\,,
\ee
where $\alpha$ is again a dimensional parameter and $G$ is the Gauss-Bonnet four dimensional topological invariant introduced in Sec. \S \ref{SecIII}.
The Friedmann-like equations read, 
\begin{equation}
16\pi \rho=6\left(H^2+\frac{k}{a^2}\right)+2\alpha H\left(3H^4+\frac{2k}{a^3}\right)
\,,
\end{equation}
\begin{equation}
16\pi (\rho+p)=-4\left(\dot H-\frac{k}{a^2}\right) -10\alpha H^3\dot H+\frac{4\alpha k}{3H a^3}\left(-\dot H+3H^2\right)\,. 
\end{equation}
Thus, by following the procedure of the previous example we find,
\begin{equation}
dE=WdV_H +T_H\left(2\pi r_H d r_H-5\pi\alpha r_H^4 H^4 dr_H\right)
+ T_H\frac{2\pi r_H k \alpha}{a^3}\left(\frac{H r_H^3 da}{a}-\frac{r_H^3 dH}{3}\right)\,.
\end{equation}
For $k=0$ this expression simply reads,
\begin{equation}
dE=WdV_H +T_H\left(2\pi r_H d r_H-5\pi\alpha d r_H\right)\,,
\end{equation}
and
\begin{equation}
S=\frac{A_H}{4}\left(1-\frac{5\alpha}{r_H}\right)\,.\label{SNPG2}
\end{equation}
As a result, in this class of models, when $k=0$, power-law corrections to the area law assume the form,
\begin{equation}
S=\frac{A_H}{4}\left(1-\gamma\frac{1}{A_H^\nu}\right)\,,
\end{equation}
where $\gamma$ is a constant and $1>\nu>0$. This corrections may emerge in dealing with the entanglement of quantum fields \cite{powerlawcorr}. 
In (\ref{SNPG1}) and (\ref{SNPG2}) we have $\nu=\frac{1}{2}$. We also observe that for $\alpha\ll \frac{1}{H}$ the corrections to the area law are negligible in both of the models.

\section{Extended mimetic gravity}

Mimetic gravity was firstly introduced by Mukhanov and Chamseddine
in Ref. \cite{Muk0} (see  also Ref. \cite{Muk1} for a minimal extension of the model). Thanks to a (singular) disformal transformation of the metric which leads to the presence of a scalar ``mimetic'' field, the theory is able to reproduce the cosmological dark matter without invoking any exotic fluid, and the solutions of GR can be recovered as a special case. However, in the original formulation of mimetic gravity,  scalar perturbations cannot propagate and some additional terms depending on the mimetic scalar field need to be added in the lagrangian. In this section we will work with an extended mimetic model \cite{mimetic}
where the field equations on FLRW space-time are at the second order.
The action of the model is given by,
\be
I =  \frac{1}{8\pi}\int_{\mathcal M} d^4 x \sqrt{-g} \left[\frac{R}{2} + \lambda\left(X-\frac{1}{2}\right)+f[\chi(\phi)] \right]+I_m\,,
\ee
where $X=-\frac{1}{2}g^{\mu \nu}\partial_\mu \phi \partial_\nu \phi$,  $\lambda$ is a Lagrange multiplier, $\phi$ is a mimetic scalar field and $I_m$ is the matter-radiation action of a perfect fluid. 
Moreover, $f(\chi)\equiv f(\chi(\phi))$ is a generic function which depends on
the higher order differential term in $\phi$, namely $\chi(\phi)=-\nabla^\mu \nabla_\mu \phi \, /3$. Thus, taking into account that on FLRW metric the Lagrange multiplier constrains the mimetic field to behave as $\phi= t$, we simply get
\begin{equation}
\chi=H
\quad\quad\rightarrow \quad\quad
f(\chi)=f(H)\,.
\end{equation}
 The Friedmann-like equations can be written as
 \be
 16\pi\rho =
 6\left(H^2+\frac{k}{a^2}\right)+f(H)-H f_H   \,,
\label{uno}
\ee
\be
16\pi(\rho+p)=
-4\dot H+
\frac{4k}{a^2}
+\frac{1}{3}\dot f_H
\,,
\label{due}
\ee
with the usual conservation law (\ref{conslaw}) for ordinary matter.
Here, $\rho\rightarrow \rho+\frac{C}{a^3}$, where $C$ is a constant.
The additional ``dark matter'' term comes from mimetic scalar field, as in the original work of Mukhanov and Chamseddine. 

From Eq. (\ref{uno}) on the apparent horizon we derive
\begin{equation}
16\pi d\rho=
-\frac{12}{r_H^3}d r_H-H d f_H\,.
\end{equation}
Thus, if we use (\ref{dE}) together with (\ref{due}) we get
\begin{equation}
dE=WdV_H+dr_H\left(-1+\frac{\dot r_H}{2H r_H}\right) +\frac{H r_H^3}{12}d f_H\left(-1+\frac{\dot r_H}{2H r_H}\right)\,.
\end{equation}
By introducing the temperature associated to the apparent horizon (\ref{TH}) we have
\begin{equation}
d E=W dV_H + 
\left( 2\pi r_H d r_H+ \frac{\pi H r_H^4}{6} d f_H\right) T_H\,.
\end{equation}
As a result, the first law brings to the identification
\begin{equation}
d S= 2\pi r_H d r_H+ \frac{\pi H r_H^4}{6} d f_H
=d\left(\pi r_H^2\right)
+\frac{\pi}{6}H\dot H \left(\frac{a^2}{H^2 a^2+k}\right)^2 f_{HH} dt\,.
\end{equation}
In fact, the area law of GR is corrected by a term depending on the mimetic scalar field through the function $f_{HH}$. This result is valid on FLRW space-time only, where for a specific model one may compute the entropy of the apparent horizon once the form of the scale factor is fixed. However, as in the previous examples, a generalization is possible 
in the flat space with $k=0$ such that
\begin{equation}
d S=
\left(2\pi r_H dr_H+\frac{\pi}{6H^3}\frac{d^2 f(H)}{d H^2}dH
\right)\,,
\label{dSfH}
\end{equation}
and the correction to the area law can be computed integrating with respect to the Hubble parameter. For example, in Born-Infield inflationary scenario considered in Ref. \cite{mimetic} we obtain (we omit some dimensional parameter),
\begin{equation}
f(\chi)=1+\frac{\chi^2}{2}-\chi\arcsin\chi-\sqrt{1-\chi^2}\,,
\end{equation}
and the entropy related to the apparent horizon results to be,
\begin{equation}
S=\frac{A_H}{4}+\frac{\pi}{12}\left[
-\ln \left[ \sqrt{\frac{H+1}{H-1}}-1\right]
+\ln\left[\sqrt{\frac{H+1}{H-1}}+1\right]
+H^2\sqrt{\frac{H+1}{H-1}}-H\sqrt{\frac{H+1}{H-1}}-\frac{2H+1}{H^2}
\right]\,.
\end{equation}
Alternatively, formula (\ref{dSfH}) can be written as
\begin{equation}
dS=2\pi r_H dr_H\left(1-\frac{f_{HH}}{12}\right)\,.
\end{equation}
As in the case of 
$F(R,G)$-gravity with second order Friedmann-like equations,
 one may recover some interesting scenarios in the context of modified entropy law. For example, it is easy to see that corrections of the type
\begin{equation}
f(\chi)=-\frac{\gamma \chi^4}{\pi}\,,
\end{equation}
with $\gamma$ dimensional  constant, lead to logarithmic-corrected entropy,
\begin{equation}
S=\frac{A_H}{4}+\gamma \ln \left(\frac{A_H}{4}\right)\,.
\end{equation}
A large variety of scenarios predict logharitmic-corrected entropy as the leading order quantum gravitational correction to the Bekenstein-Hawking entropy \cite{logcorr1, logcorr2, logcorr3}, mainly motivated by conformal anomaly \cite{conf} and 
quantum tunneling \cite{tun1, tun2} (see also Ref. \cite{Xiao} and references therein).

\section{Mimetic Horndenski inspired gravity \label{6}}

Horndeski gravity is a class of scalar tensor theories
where the scalar field interacts with gravity and
the field equations are at the second order like in General Relativity \cite{Horn}. In literature these theories have been well studied in different scenarios \cite{Hornworks1, Hornworks2,Hornworks3, Hornworks4,Hornworks5, Hornworks6}. Here, we remain in the context of mimetic gravity where the scalar field plays the role of dark matter, and we recall a model \cite{Vagnozzi0} where
 mimetic gravity action is implemented with higher-order terms that break the Horndeski structure of the lagrangian but still preserve second-order field equations on FLRW background. In this model scalar perturbations can propagate and gravitational wave speed is close enough to the speed of light ensuring the agreement with last cosmological data \cite{Vagnozzi1}. 

The action of the model is given by, 
\begin{equation}
\frac{1}{8\pi}\int_{\mathcal M} d^4\sqrt{-g}
\left[\frac{R}{2}(1+2a X)-\frac{c}{4}\left(\Box \phi\right)^2+\frac{b}{4}\left(\nabla_\mu\nabla_\nu\phi\right)^2-\frac{\lambda}{4}\left(2X+1\right)\right]+I_m\,,
\end{equation}
where, again, $X=-\frac{1}{2}g^{\mu \nu}\partial_\mu \phi \partial_\nu \phi$,  $\lambda$ is a Lagrange multiplier, $\phi$ is the mimetic scalar field and $I_m$ is the matter-radiation action of a perfect fluid. When $b=c=4a$ we recover a Horndenski mimetic model. Due to the identification of $\phi$ with the cosmological time, $\phi=t$, the equations of motion on FLRW space-time are extremely simple and read,
\begin{equation}
16\pi\rho=\left(\frac{4-b+3c-4a}{4}\right) 6\left(H^2+\frac{k}{a^2}\right)\,,
\end{equation}
\begin{equation}
16\pi(\rho+p)=\left(\frac{4-b+3c-4a}{4}\right)
4\left(
-\dot H+\frac{k}{a^2}
\right)\,.
\end{equation}
The mimetic scalar field contributes to the energy density as $\rho\rightarrow \rho+\frac{C}{a^3}$, where $C$ is a constant. Thus, when $a=b=c=0$ we recover the Friedmann equations of GR with the additional contribution of dark matter. 

Now the first law is given by,
\begin{equation}
dE=W dV_H+ \left(\frac{4-b+3c-4a}{4}\right)(2\pi r_H dr_H)T_H\,,
\end{equation}
and the entropy of the apparent horizon reads,
\begin{equation}
S=\left(4-b+3c-4a\right)\frac{A_H}{16}\,,
\end{equation}
and is proportional to the area of the horizon as in GR. In Ref. \cite{Vagnozzi1} it has been found that in light of last cosmological data the parameters $b$ and $c$ must be extremely close to zero, while the parameter $a$ should be $0\leq a<1$ in order to avoid ghost instabilities. It means that by taking into account the observational constraints 
this mimetic model predicts a (positive)
entropy in the form,
\begin{equation}
S\simeq (1-a)\frac{A_H}{4}\,.
\end{equation}
Thus, the value of the parameter $a$ 
affects the deviation of entropy result with respect to the area law of GR.

\section{Conclusions}

In this paper, we have investigated the first law of thermodynamics on (non flat) FLRW space-time in different theories of modified gravity. The thermodynamical proprieties of the black hole horizon are extended to the apparent horizon of FLRW space-time where the temperature is derived from the metric through the surface gravity.
Furthermore, one expects that the first law associated to the apparent horizon of FLRW space-time is satisfied and the field equations of a gravitational theory should be  consistent with its formulation. 

In FLRW universe a matter-radiation source of perfect fluid 
is present and the energy and the working term of the first law can be described by energy density and pressure of the fluid. The formalism has been tested in General Relativity, Gauss-Bonnet and Lovelock gravity, where one recovers the entropy area law.

In modified gravity the entropy associated to apparent horizon does not satisfy the area law. Several alternatives to Bekenstein-Hawking entropy mainly motivated by quantum effects
already exist, and in principle one may try to reconstruct the corresponding gravitational field equations by starting from the first law. On the other hand, the issue to formulate the gravitational theory which leads to a given entropy is much more complicate, due to the fact that when one works with a modified framework the field equations are not longer at the second order and we lose many of the proprieties of GR. 

For this reasons, it is interesting to see what it is possible to argue in those (higher derivative) theories where the field equations on FLRW background remain at the second order. In our examples, at least for the flat spatial case, we found a formula for entropy in terms of the radius of the apparent horizon only, which allows to reconstruct the gravitational lagrangians associated to some given entropy laws.

First of all, we derived the first law of thermodynamics of non-flat FLRW universe in the framework of $F(R, G)$-gravity, were some additional terms to the Wald entropy emerge and in the case of $F(R)$-gravity we recover the result of Ref. \cite{SuperCai}. Then, we investigated a special class of $F(R, G)$-models where the field equations on FLRW space-time are at the second order and we showed that in the flat space it is possible to derive a formula for the entropy which is independent of the explicit form of the scale factor. 
Similar results are found for a class of non-polynomial models depending on the covariant derivatives of the Ricci scalar and the Gauss-Bonnet where the field equations on FLRW space-time are at the second order. Some attempts to recover specific modified entropy laws are presented.

In the second part of the paper we considered two extended mimetic gravity models where a scalar field plays the role of dark matter on FLRW background. The dark matter contribution enters in the first law by increasing the total energy density of matter-radiation contents of the universe, while the terms added to pure mimetic gravity modify the entropy area law. We remark that the viability of the model presented in \S \ref{6} has been investigated in the light of cosmological data in Ref. \cite{Vagnozzi1}.

\section*{Appendix}

The entropy associated to $F(R,G)$-gravity can
be calculated via Wald's method~\cite{Wald}.
The explicit calculation of entropy $S_W$ is given by the formula~\cite{Visser:1993nu, FaraoniEntropy},
\begin{equation}
S_W = - \frac{1}{8}
\oint_{\tiny{\begin{array}{cc}
r=  r_H \\
t = \mbox{const}
\end{array}}}
\left. \left(\frac{\delta F(R, G)}{\delta R_{\mu\nu\xi \sigma}}\right)\right.\,
e_{\mu \nu} e_{\sigma \xi}r\, d\theta\, d\phi\,.
\label{Wald}
\end{equation}
The antisymmetric variable
$e_{\mu\nu}=-e_{\nu\mu}$
is the binormal vector to the (bifurcate) horizon and
it is normalized so that $e_{\mu\nu}e^{\mu\nu}=-2$. Thus, by using the FLRW metric, it turns out to be 
\begin{equation}
\epsilon_{\mu\nu}=
\sqrt{\frac{a(t)^2}{(1+k r^2)}}
(\delta^0_{\mu}\delta^1_{\nu}-\delta^1_{\mu}\delta^0_{\nu})\,,
\end{equation}
$\delta^i_j$ being the Kronecker delta.

By taking
the variation of $F(R, G)$ with respect to $R_{\mu\nu \xi \sigma}$ as if $R_{\mu\nu \xi \sigma}$ and the metric $g_{\mu\nu}$ are independent, formula (\ref{Wald}) leads to
\begin{eqnarray}
S_W &=& -\frac{1}{2}A_H\,
\left(\frac{a^2}{1+k r^2}\right)
\,\left(\frac{\delta F(R, G)}{\delta R_{0 1 0
1}}\right)\Big\vert_H\nonumber\\
&=&
-\frac{1}{2}A_H\,
\left(\frac{a^2}{1+k r^2}\right)
\left(F_R\frac{\delta R}{\delta R_{0 1 0 1}}+ F_G\frac{\delta G}{\delta R_{0 1 0 1}}\right)\Big\vert_H\label{waldbis}\,,
\end{eqnarray}
with $A_H=4\pi r_H^2$. Since
\begin{equation}
\frac{\delta R}{\delta R_{\mu \nu \alpha \beta }}=
\frac{1}{2}\left(g^{\alpha \mu}g^{\nu \beta}-g^{\nu \alpha}g^{\mu \beta}  \right)\,,
\end{equation}
\begin{equation}
\frac{\delta G}{\delta R_{\mu\nu\xi\sigma}}=
\left[ 2 R^{\mu\nu\xi\sigma} -2 (g^{\mu \xi} R^{\nu \sigma} + g^{\nu
\sigma} R^{\mu \xi} - g^{\mu \sigma} R^{\nu \xi} - g^{\nu \xi} R^{\mu \sigma})+
(g^{\mu \xi} g^{\nu \sigma} - g^{\mu \sigma} g^{\nu \xi}) R \right]\,,
\label{variation}
\end{equation}
one obtains,
\begin{equation}
S_W=\frac{A_H}{4 }\left(F_R+ F_G\left(\frac{4}{r_H^2}\right)\right)\,.
\end{equation} 
In General Relativity $F(R,G)=R$ and one recovers the usual area law, namely $S_W= A_H/4$. The result is in agreement with the spherical case of Ref. \cite{mioFRG}, where static spherical symmetric black hole solutions and the associated Wald entropy in $F(R, G)$-gravity are investigated.

\end{document}